\documentstyle[twoside,11pt,aaspp4]{article}
\tighten
\eqsecnum
\received{}
\accepted{}
\journalid{}{}
\articleid{}{}
\slugcomment{To appear in {\it The Astronomical Journal}}
 
\begin{document}

\title{The Lithium-Rotation Correlation in the Pleiades Revisited}

\author{Jeremy R. King}
\affil{Department of Physics, University of Nevada, Las
Vegas\altaffilmark{1}, 4505 S. Maryland Parkway, Las Vegas, NV
89154-4002\\ and \\ 
Space Telescope Science Institute, 3700 San Martin Drive,
Baltimore, MD 21218\\ email: jking@physics.unlv.edu}

\author{Anita Krishnamurthi}
\affil{JILA, University of Colorado and National Institute of
Standards and Technology, Campus Box 440, Boulder, CO 80309-0440\\
email: anitak@casa.colorado.edu}

\author{Marc H. Pinsonneault}
\affil{Department of Astronomy, The Ohio State University, 140 W. 18th
Ave., Columbus, OH 43210\\ email: pinsono@astronomy.ohio-state.edu} 

\altaffiltext{1}{Current address}

\begin{abstract}

The dispersion in lithium abundance at fixed effective temperature in
young cool stars like the Pleiades has proved a difficult challenge for
stellar evolution theory.  We propose that Li abundances relative to a
mean temperature trend, rather than the absolute abundances, should be
used to analyze the spread in abundance.  We present evidence that the
dispersion in Li equivalent widths at fixed color in cool single
Pleiades stars is at least partially caused by stellar atmosphere
effects (most likely departures from ionization predictions of model
{\it photospheres\/}) rather than being completely explained by genuine
abundance differences.  We find that effective temperature estimates
from different colors yield systematically different values for active
stars.  There is also a significant correlation between stellar activity and
Li excess, but not a one-to-one mapping between unprojected stellar
rotation (from photometric periods) and Li excess.  Thus, it is unlikely
that rotation is the main cause for the dispersion in the Li abundances.
Finally, there is a signficant correlation between detrended Li excess and
potassium excess but not calcium-- perhaps supporting incomplete
radiative transfer calculations (and overionization effects in
particular) as an important source of the Li scatter.  Other mechanisms,
such as very small metallicity variations and magnetic fields, which
influence PMS Li burning may also play a role.  Finally, we find no
statistical evidence for a decrease in dispersion in the coolest
Pleiades stars, contrary to some previous work.

\end{abstract}
 
\keywords{open clusters and associations: individual (Pleiades) --- 
stars: abundances, activity, atmospheres, interiors, late-type, rotation} 

\section{Introduction}

Predictions are made by standard stellar models (e.g., \cite{bahcall88})
about the surface abundances of elements in stars.  However, there are
indications that such models are incomplete.  A case in point is the
surface abundance of the element lithium (Li) in low mass stars, which
is observed to decrease with time.

The solar meteoritic value for the Li abundance is 3.31$\pm$0.04
(\cite{anders89}).  A study of Li abundances in young, pre-main-sequence
(PMS) T Tauri Stars (TTS) suggests a value of log N(Li) = 3.2$\pm$0.3
(\cite{magazzu92}), consistent with the meteoritic value.  While this is
one indicator of the initial Li abundance, TTS abundance determinations
are beset by complications due to their youth, such as uncertain $T_{\rm
eff}$ estimates and the presence of a circumstellar accretion disk.  The
study of stars in open clusters of different ages, like $\alpha$ Persei
(50 Myr) and the Pleiades (70--100 Myr), shows that there is nearly a
uniform Li abundance of 3.2 for high mass stars (${\sim}7000$ K).

It is known that surface Li depletion takes place during the PMS
evolution of low mass stars due to Li burning via ($p$, $\alpha$)
reactions at low temperatures of $T{\ga}2.6{\times}10^{6}$ K.  Surface
depletion can occur in standard models through convective mixing if the
base of the convection zone is hot enough to burn Li (\cite{bode65},
\cite{pinsono97}).  Because PMS stars have deep convection zones, they
burn Li during the PMS.  As the depth of the convection zone is a
function of mass (increasing with lower mass) Li is depleted on the main
sequence only in lower mass stars (${\leq}0.9$ M$_{\odot}$).  However,
the Pleiades evinces a large dispersion in surface Li abundance at a
given color for $T_{\rm eff}{\la}5500$ K (e.g., \cite{s93b}).  Standard
stellar models are unable to reproduce this dispersion.  Furthermore,
open cluster observations indicate some depletion is observed on the
main sequence as well, which is in conflict with standard standard
models.

Because these models cannot fully explain the observed depletion
patterns, additional mixing mechanisms seem necessary.  Rotation
provides one driving mechanism for such non-convective mixing, through
meridional circulation (\cite{tassoul78}, \cite{zahn92}) and
instabilities caused by differential rotation (\cite{zahn83}).  Hence,
rotation in stars has received much scrutiny as a possible agent of Li
depletion and of the observed scatter in open cluster Li abundances at a
given mass.  Models which include rotational mixing (\cite{pkd90}) are
able to predict the dispersion seen in older systems, but not at young
ages like that of the Pleiades (\cite{cpd95}).  The study of Li
abundances is a rich and vast field, and there have been several efforts
to study the correlation of surface Li abundances with rotation using
stars in open clusters.  Here, we concentrate on the connection between
surface Li abundance and rotation using data in the young Pleiades
cluster.

Because the Pleiades Li scatter is such a difficult obstacle in our
understanding of early stellar evolution, a historical summary seems in
order.  Butler et al.~(1987) studied a sample of 11 K-stars in the
Pleiades and determined that four rapid rotators had higher Li
abundances than four slow rotators.  They believed this consistent with
the evolutionary picture that on arrival on the main sequence, stars had
high rotation rates and high Li abundances (i.e., they arrived on the
main sequence before there was time for rotational braking or Li
depletion).  As the star spun down, the Li abundance decreased as well.
Hence, they concluded that the faster rotators were younger than the
slower, hence less depleted.

A study of the distribution of rotational velocities of low-mass stars
in the Pleiades by Stauffer \& Hartmann (1987) revealed that there was a
wide range of rotation velocities in the Pleiades K and M dwarfs.  They
showed that the distribution of rotation velocities in the Pleiades
could be reproduced quite well invoking angular momentum loss, without
having to resort to a large age spread which is also in conflict with
the narrow main-sequence seen among the low-mass Pleiades stars.

Soderblom et al.~(1993b) carried out an extensive study of Li abundances
in the Pleiades.  They considered several possible explanations for the
dispersion in the observed abundances, including observational errors
and the effect of starspots.  They concluded that the spread in Li
abundances seen was real and not an artifact of other physical
conditions.  They found that the Li abundance was correlated well with
both rotation and chromospheric activity, and speculated that rapid
rotation was somehow able to preserve Li in stars.  While they found
some low $v \sin i$ systems with high Li abundances, it was possible
that these stars are faster rotators simply seen at low inclination.

Balachandran et al.~(1988) studied a sample of stars in the younger
$\alpha$ Persei cluster (50 Myr) and concurred with the picture of
Li-poor stars as slow rotators.  However, a comparison of the Pleiades
to $\alpha$ Per by Soderblom et al.~(1993b) showed that while most stars
had similar abundances, a significant number of stars in $\alpha$ Per
had abundances that were less than that in the Pleiades by 1 dex or
more.  This was difficult to understand until Balachandran et al.~(1996)
published a corrected list of Li abundances that culled all non-members
from the sample, bringing consistency to the Pleiades and $\alpha$ Per
abundances.

Garcia Lopez et al.~(1991a,b) added seven stars to the Butler et
al.~sample in the range $4500{\ga}T_{\rm eff}{\ga}5500$ K and asserted a
clear connection between Li abundance and $v \sin i$.  They also found
that the correlation breaks down for temperatures cooler than 4500 K.
In a subsequent paper in 1994, they enlarged their sample and further
studied the correlation, concluding that their earlier assertion was
correct -- there were no rapid rotators with low Li abundances and there
was a clear relationship between log N(Li) and $v \sin i$.  They did
note three stars (\ion{H}{2} 320, 380 and 1124) having low $v \sin i$
values and Li abundances comparable to those of the rapid rotators as
counterexamples, but speculated these objects were rapid rotators seen
at low inclination angles.  It is to be noted that, when determining
mean abundances for their rapid and slow rotator populations, they
included the 3 stars with low $v \sin i$ and high Li abundances in their
rapid rotator sample.  While this does not change the qualitative result
they obtained, it does affect the magnitude of the difference in
abundance between the two populations.  It also demonstrates that there
is a range of abundances for slow rotators.

Jones et al.~(1996a) derived Li abundances and rotation velocities for
15 late-K Pleiades dwarfs, and also found that the correlation between
Li abundance and rapid rotation breaks down for cooler stars ($T_{\rm
eff}{\la}4400$ K).  Jones et al.~(1997) determined rotational velocities
and Li abundances in the 250 Myr old cluster M34, intermediate in age
between the young Pleiades (70--100 Myr) and the Hyades (500 Myr).  They
concluded that the Li depletion and rotation velocities were in between
the Pleiades and Hyades values, and that the pattern seen in these
clusters suggested an evolutionary sequence for angular momentum loss
and Li depletion.

One could speculate that some of the Li dispersion in the Pleiades and
$\alpha$ Per may be due to NLTE effects and unknown effects of stellar
activity on the \ion{Li}{1} line formation (\cite{houd95};
\cite{russ96}; but see Soderblom et al.~1993b).  The structural effects
of rotation might also be responsible for the Li depletion pattern.
Martin \& Claret (1996) included this ingredient in their models for
masses of 0.7 and 0.8 M$_{\odot}$ and were able to produce enhanced Li
abundances for stars with high initial angular momenta as a result of
less effective PMS Li destruction in rapid rotators (i.e., their models
imply initially rapid rotators will have high Li abundances relative to
the other stars at the same $T_{\rm eff}$ at young ages).  Angular
momentum loss as well as rotationally-induced mixing could affect these
models.

The questions we consider here are if there truly is a correlation
between Li abundance and rotation rate in the Pleiades, what the nature
of the correlation is, and if not, what might explain the abundance
scatter.  We start with a careful sample selection for our analysis, and
examine various possible causes that might contribute to the dispersion
(including errors in abundance determination).  We then proceed to an
analysis of the Li-rotation correlation and explore other possible
correlations that might be masquerading as a Li-rotation correlation.

\section{Pleiades Lithium Abundances}

\subsection{Sample Selection and Definition}

Lithium abundances were derived from the datasets in the studies of
\markcite{s93a}Soderblom {\it et al.\/}~(1993a; S93a) and
\markcite{j96}Jones {\it et al.\/}~(1996).  The two studies were merged
to form our starting sample with the latter data preferred in cases of
overlap.  Secondary stellar companions can affect photometric colors
from which $T_{\rm eff}$ values are derived, activity levels, measured
line strengths, and activity levels deduced spectroscopically.  Thus, in
order to look at the intrinsic Pleiades Li abundance dispersion
unrelated (directly or indirectly) to the presence of a stellar
companion, binary systems were excised from our sample.  Cluster and
interloping field binaries identified by \markcite{m92}Mermilliod {\it
et al.\/}~(1992) and \markcite{b97}Bouvier {\it et al.\/}~(1997) were
removed from the starting sample.  Two spectroscopic binaries not in
these lists, but identified as such by S93a, were also removed.

The color-magnitude diagram of this refined sample was then inspected to
photometrically identify binaries using the dereddened $BVI$ photometry
described by \markcite{P98}Pinsonneault {\it et al.\/}~(1998).  We found
\ion{H}{2} 739 to be an obviously overluminous (or overly red) outliar
in the $V$ vs.~$B-V$, $V$ vs.~$V-I$, and $I$ vs.~$V-I$ diagrams, and
eliminated it from the refined sample.  Finally, all stars with upper
limits on the ${\lambda}6707$ \ion{Li}{1} line's equivalent width were
eliminated.  These upper limits, as censored data, complicate the
ensuing statistical analysis.  These stars are also the very
hottest and very coolest in the sample.  Their photometrically-inferred
$T_{\rm eff}$ values and model atmospheres may be slightly more
uncertain than the other objects in the sample.  Their elimination
simplifies the analysis and reduces possible additional sources of 
uncertainty. 

This final sample of 76 Pleiades stars is listed in the first column of
Table 1.  The extinction-corrected $V$ magnitude and reddening-corrected
$(B-V)$ and $(V-I)$ colors are given in the second, third, and fourth
columns.  The color-magnitude diagram of these stars evinces a tight
main sequence, and is shown in Figure 1 (open circles) with a 100 Myr
isochrone described in Pinsonneault {\it et al.\/}~(1998) and assuming a
distance modulus of 5.63 (Pinsonneault {\it et al.\/}~1998).  The only
possibly discrepant outliars remaining are: {\ }a) \ion{H}{2} 686, which
appears overluminous in the V vs.~$B-V$ plane, but not the $V-I$ plane
{\ }b) \ion{H}{2} 676, which appears underluminous (or too blue) in the
$V-I$ plane and perhaps $B-V$ also, and {\ }c) \ion{H}{2} 2034, which
appears underluminous (or too blue) in the $B-V$ plane, but not in
$V-I$.  There is no convincing evidence that these slight discrepancies
are related to binarity.  Rather, they may be due to relatively large
photometric errors in one passband or to physical effects ({\it e.g.,\
\/}increased red flux from spots) unrelated to binarity.  Stars rejected
as binaries are plotted as filled triangles; their general propensity to
reside above the main sequence is evident.
\marginpar{Tab.~1}
\marginpar{Fig.~1} 

\subsection{Stellar $T_{\rm eff}$ and Activity Measures} 

S93a and Jones {\it et al.\/}~(1996) provide photometric $T_{\rm eff}$
estimates for all of the stars in Table 1.  We re-examine these for
comparison and because of concern that chromospheric activity or
starspots might affect the colors of young stars.  We calculated $T_{\rm
eff}$ using our $(B-V)_{\rm o}$ values and the relation from Soderblom
et al. (1993b, equation 3): $T_{\rm eff}=1808(B-V)_{\rm
o}^2-6103(B-V)_{\rm o}+8899$.  Temperatures were also derived from our
$(V-I)_{\rm o}$ colors using the relation from Randich et al. (1997):
$T_{\rm eff}=9900-8598(V-I)_{\rm o}+4246(V-I)_{\rm o}^2-755(V-I)_{\rm
o}^3$.  Both of these relations are based on the data from Bessell
(1979), and should provide self-consistent temperatures given
self-consistent photospheric colors.  Columns 5, 6 and 9 give the
$T_{\rm eff}$ values of S93a, and those derived here from $(B-V)$ and
$(V-I)$.

We adopt the H${\alpha}$- and \ion{Ca}{2} infrared triplet-based
chromospheric emission measurements from S93a as stellar activity
indicators.  These are the ratio of the flux (relative to an inactive
star of similar color) in the H${\alpha}$ and \ion{Ca}{2} lines relative
to the total stellar bolometric flux.  Given canonical views of a
relation between stellar mass and chromospheric emission on the main
sequence, it is also of interest to measure the residual H$\alpha$ and
\ion{Ca}{2} flux ratios.  That is, we wish to detrend the general
relation between stellar mass and activity such that activity
differences unrelated to large-scale mass differences can be quantified.
This was done by fitting the H$\alpha$ and \ion{Ca}{2} flux ratios as a
function of $(V-I)$ color temperature with a linear
relation\footnote{Quantitative comparison of the resulting ${\chi}^2$
values indicated that fits with higher order functions did not yield
statistically improved descriptions of the flux ratio-color relations.},
and subtracting this fitted flux ratio (computed at a given $V-I$) from
the measured flux ratio of each star.  The relation for the fitted
H$\alpha$ flux ratio (used below) was found to be $\log R_{{\rm
H}{\alpha},{\rm fit}}=(-0.00044742{\times}T_{\rm eff}(V-I))-2.12515$.
The relation for the \ion{Ca}{2} flux ratio was found to be $\log R_{\rm
CaII,fit}=(-0.00021017{\times}T_{\rm eff}(V-I))-3.50280$.

We find strong evidence that our $T_{\rm eff}$ values (hence, assuming
self-consistency of the color-$T_{\rm eff}$ relations, the photometric
colors) are affected by activity level.  Figure 2 shows the difference
between the $(B-V)$- and $(V-I)$-based $T_{\rm eff}$ values versus the
H$\alpha$ flux ratios (top panel) and the mass-independent residual
H$\alpha$ flux ratios (bottom panel).  A relation is seen in both
panels, such that the lowest $T_{\rm eff}$ differences are seen
predominantly for the lowest flux ratios while the largest $T_{\rm eff}$
differences are seen predominantly for stars having the largest flux
ratios.  The ordinary linear correlation coefficients are significant
above the 99.9\% confidence levels for both panels.
\marginpar{Fig.~2}

The binary stars (filled triangles) behave similarly to the single stars
in both panels; on average, though, the binaries exhibit larger $T_{\rm
eff}$ residuals than the single stars.  This systematic offset likely
reflects the additional influence of fainter (hence cooler and redder)
companions on the photometric colors.  If this interpretation is
correct, it could suggest that the handful of inactive single stars with
significant $T_{\rm eff}$ residuals in the upper left portion of both
panels are unrecognized binaries.

Significant differences between the $(B-V)$- and $(V-I)$-based $T_{\rm
eff}$ estimates, the slight propensity for $(B-V)$ to yield larger
temperatures, and the association of these properties with stellar
activity seems to be a common property of young stars noted and
discussed by others ({\it e.g.,\ \/}\markcite{g94}Garcia Lopez {\it et
al.\/}~1994; \markcite{r97}Randich {\it et al.\/}~1997;
\markcite{k98}King 1998; \markcite{s99}Soderblom {\it et al.\/}~1999).
Explanations for these observed properties in young stars are at least
twofold: {\ }a) increased $B$-band flux due to boundary layer emission
associated with a circumstellar disk (presumably not applicable for our
near-ZAMS Pleiads), and {\ }b) increased $I$-band flux due to the
presence of cool spots.  It is straightforward to associate increased
prevalence and surface coverage of spots with increasing activity, and
H$\alpha$ emission (used here to quantify activity) has been associated
with accretion of circumstellar material in young stars.  Given that the
Pleiades age (${\sim}100$ Myr) is an order of magnitude larger than
inferred disk lifetimes for solar-type stars (\markcite{s90}Skrutskie
{\it et al.\/}~1990), spots are the more likely cause of the temperature
difference-activity relation in our Pleiades sample.

In their recent study of the effects of activity on Pleiades Li
abundances, \markcite{SBR}Stuik {\it et al.\/}~(1997) find that
activity-- specifically the presence of spots and plages-- may
significantly alter photospheric colors.  Indeed, they suggest that the
resulting changes in color may be a more dominant contributor to the
Pleiades Li spread than line strength differences.  Additionally, they
find that such color variations are both surprising and complex.  Their
empirical solar-based activity models indicated that {\it both\ \/}spots
and plages lead to increased $(B-V)$ colors; in contrast, their
``best-effort'' theoretical stellar models indicate a decrease in
$(B-V)$.  Sorting out which (if either) set of models are appropriate
for specific Pleiads (in addition to other empirical details such as
specific spot/plage coverage and ratio) might be further illuminating.
Because the direction of changes in $(V-I)$ also flips in their models,
Stuik et al.~(1997) note that spot/plage-related changes in color may
not be ideally identified in two-color plots (e.g., Soderblom {\it et
al.\/}~1993).

\subsection{Lithium Abundance Determinations and Detrending}

Li abundances for all our Pleiads were determined from the measured 
${\lambda}6707$ line strengths and our preferred $T_{\rm eff}$ value.  A
blending complex lies some 0.4 {\AA} blueward of the Li doublet.  In our
stars, the typical contribution of these blending features is ${\sim}10$
m{\AA}, which is significantly smaller than typical Li line strength of
${\sim}100$ m{\AA}.  The blending contribution was 
subtracted\footnote{Deblending corrections were not applied to any stars 
taken from Jones et al.~(1996) following their claim that instrumental 
resolution was sufficient to separate the Li line and blending complex.
While it is not clear to us that this is true given that some of their 
objects have appreciable rotation (see their Figure 1), it does not affect
the present results inasmuch as the Jones et al.~rapid rotators have 
Li line strengths significantly larger than those expected of the 
blending complex.} following the approach of S93a, who parameterized 
the contaminating line strength as a function of $(B-V)$ color.  Here, 
we recast this parameterization as a function of $T_{\rm eff}$ so that 
differences in our $(B-V)$- and $(V-I)$-based temperatures were consistently 
accounted for in the analysis.

Given the $T_{\rm eff}$ values and the corrected Li line strengths,
abundances were calculated using Table 2 from S93a.  This was done by
fitting a surface map of the equivalent width-temperature-abundance grid
of S93a using high order polynomials.  The internal interpolation
accuracy is generally a few thousandths of a dex\footnote{The two
hottest single stars have $T_{\rm eff}$ values significantly outside the
curve of growth grids provided by S93a.  Extrapolation to these
temperatures with high order polynomials leads to errantly low
abundances by a few tenths of a dex.  To the extent that we are mainly
interested in the cooler Pleiads and that we are only interested in the
differential star-to-star Li abundances (i.e., large scale abundance
morphology with $T_{\rm eff}$ is removed later), these known errors are
unimportant for the present analysis.  In the case of these two stars
(\ion{H}{2} 133 and 470), we simply caution those who would use our
absolute abundances, and also note that the few much smaller
extrapolations to lower $T_{\rm eff}$ outside the S93a grid are not
believed to be affected by any substantial amount.}.  Columns 8 and 11
of Table 1 give the derived Li abundances\footnote{by number, relative
to hydrogen, on the usual scale with log $N$(H)$=12.$} for our
$(B-V)$-based $T_{\rm eff}$, and for our $(V-I)$-based $T_{\rm eff}$.
 
At the Pleiades age, PMS Li burning has significantly depleted the
initial photospheric Li content of many of our stars.  Moreover, this
PMS depletion is a sensitive function of mass (or $T_{\rm eff}$) with
less massive stars having depleted more Li due to deeper convection
zones and longer PMS evolutionary timescales.  In examining star-to-star
Li abundance differences connected with parameters such as rotation or
activity, we must remove this general large-scale trend in the abundance
vs.~$T_{\rm eff}$ plane.

The procedure is illustrated in Figures 3 and 4, which shows the
Pleiades Li abundances versus our $(B-V)$- and $(V-I)$-based $T_{\rm
eff}$.  The familiar and large (3 orders of magnitude) abundance
depletion over 2000 K of $T_{\rm eff}$ is seen in Figure 3.  The mean
trend is shown by the dashed line, which is a fourth order Legendre
polynomial fitted to the single star data after rejecting
${\pm}3{\sigma}$ outliars.  Fourth order fits were also conducted for
the data based on the S93a temperatures and our $(B-V)$-based values.
These fits provide a mean fiducial Li abundance at any given $T_{\rm
eff}$ to which observed abundances (calculated assuming the same source
of $T_{\rm eff}$) can be compared to infer and measure a relative Li
``enhancement'' or ``depletion'' for each star as shown in Figure 4.
For $(V-I)$-based temperatures, the approximation to the fit shown in
Figure 3 is given by: $\log N({\rm Li})_{\rm
fit}=-10.8602+(2.9785{\times}10^{-3}{\times}T)+(1.1736{\times}10^{-7}
{\times}T^2)-(3.8181{\times}10^{-11}{\times}T^3)$.  For $(B-V)$-based
temperatures, the approximation to the fit shown in Figure 3 is given
by: $\log N({\rm Li})_{\rm
fit}=-10.9959+(2.8360{\times}10^{-3}{\times}T)+(1.7354{\times}10^{-7}
{\times}T^2)-(4.2744{\times}10^{-11}{\times}T^3)$
\marginpar{Fig.~3}
\marginpar{Fig.~4}

\subsection{Errors} 

Uncertainties in the Li abundances were estimated from those in $T_{\rm
eff}$ and equivalent width.  Here, we are only interested in the
internal errors which affect the star-to-star Li abundances.  A measure
of the internal uncertainties in the $T_{\rm eff}$ estimates is provided
by the estimates from the $(B-V)$ and $(V-I)$ colors.  For the single
stars, the difference in the two color-temperatures exhibits a per star
standard deviation of 108 K.  Assuming equal contributions from both
$(B-V)$ and $(V-I)$, this suggests an internal error of ${\pm}76$ K in
the $T_{\rm eff}$ of any one Pleiad derived from any one color.  This
uncertainty was translated to a Li abundance error by re-deriving
abundances with $T_{\rm eff}$ departures of this size.  The adoption of
identical errors in (B-V) and (V-I)-based Teff values is a simplifying
assumption (though one likely true within a few tens of K). Inasmuch as
our conclusions are the same using either the (B-V) or (V-I) colors, it
is not a critical one for this work.

The other significant source of uncertainty is in the Li line
measurements.  For S93a line strengths (the majority of our sample),
uncertainties come from their own quality measures: a (${\pm}12$ m\AA),
b (${\pm}18$ m\AA), c (${\pm}25$ m\AA), and d (${\pm}40$ m\AA).  Jones
{\it et al.\/}~(1996) state that their uncertainties range from 5-20
m{\AA} and depend largely on projected rotational velocity.  Assuming
this range and the stars' $v sin i$ values, we have adopted the
reasonable values shown in column 12 of Table 1.  Given S93a's note that
the equivalent widths of \markcite{b87} Butler {\it et al.\/}~(1987) may
have to be regarded with caution, and the typical expected uncertainties
from Poisson noise expected from their S/N and resolution, we have
assigned an uncertainty of ${\pm}30$ m{\AA} in these line strengths.
Based on the S/N, resolution, and $v sin i$ values of the observations
from \markcite{b88}Boesgaard {\it et al.}\/~(1988), we have adopted a
conservative uncertainty of ${\pm}5$ m{\AA} for their data.  In a
similar fashion, we assigned uncertainties of ${\pm}25$ m{\AA} to the
equivalent widths from \markcite{p87}Pilachowski {\it et al.\/}~(1987).
The line strength uncertainties were translated to Li uncertainties by
re-deriving abundances with the adopted equivalent width departures.

Final Li abundance uncertainties, shown in Figures 3 and 4, are
calculated by summing the two errors in quadrature, and listed in
columns 13 and 14 of Table 1.  We emphasize that for the purpose of
looking at the star-to-star Li abundance differences in cool Pleiads,
the effects of $T_{\rm eff}$ errors are minimized.  This is because the
movement of a star in the $T_{\rm eff}$-Li plane due to $T_{\rm eff}$
errors is very nearly along the cool star depletion trend for $T_{\rm
eff}{\la}5800$ K.  To take into account this correlation in looking at
the {\it differential\ \/}Li abundances (i.e., the actual values versus
an expected value from a fitted trend to the data), the abundance errors
due to departures in $T_{\rm eff}$ were combined with the slope of the
fitted Li versus $T_{\rm eff}$ trend at the $T_{\rm eff}$ of each star.
The total uncertainties in the differential Li abundances are given in
the final two columns of Table 1.

\subsubsection{Li Abundance Scatter}

Large scatter in the star-to-star Li abundances is apparent in Figures 3
and 4.  Comparison of the observed scatter with that expected from the
estimated uncertainties indicates that the spread is statistically
significant.  The presence of real global scatter was considered by
comparing the variance of the differential Li abundances with that based
on the uncertainties given in Table 1.  The sizable reduced chi-squared
statistic (${\chi}^{2}_{\nu}=12.78$, ${\nu}=72$) indicates probabilities
of the observed variance ($s({\rm Li})^2{\sim}0.13$ dex$^2$) occurring
by chance are infinitesimal.

Additional analysis was carried out by binning in $T_{\rm eff}$.  For
both the $(B-V)$ and $(V-I)$ based results, we broke the data up into 5
$T_{\rm eff}$ ranges following natural breaks in the estimated $T_{\rm
eff}$ values which yielded comparable sample sizes (10-15 stars) in each
bin.  The results for both the $(B-V)$ and $(V-I)$ data are similar.

We find that stars in the hottest bin (bin `E': 6172-6984 K and
6107-6928 K for $B-V$ and $V-I$) exhibit a variance that is larger than
the expected value at only the 94\% confidence level.  The stars in the
adjacent cooler bin (bin `D': 5567-6048 K and 5521-6021 K for $B-V$ and
$V-I$) exhibit a variance in the Li abundances significantly larger than
expected from the uncertainties at the ${\ga}99.93$\% confidence level.
The differential Li abundances in the remaining three cooler bins (bin
`C': 4899-5477 K and 4996-5452 K; bin `B': 4507-4815 K and 4542-4746 K;
bin `A': 3955-4332 K and 3867-4343 K) all show observed variances
significant at considerably higher confidence levels.

An important claim by Jones {\it et al.\/}~(1996) in their study of
Pleiades Li abundances is a progressive decline in the dispersion of the
Li abundances as one proceeds from the late G dwarfs, to the
early-to-mid K dwarfs, and finally to the later K dwarfs.  We find,
however, that quantitative analysis fails to provide firm support for
such a conclusion.  F-tests of the observed variances indicate that
differences of the differential Li abundance dispersions in our cooler
three bins are statistically indistinguishable.  The differences between
the bin B and bin A stars' variances are significant at only the 71.5\%
and 78.0\% confidence levels for the $B-V$ and $V-I$ datasets.
Differences between the bin C and bin B stars' variances are different
at only the 72.7\% and 75.0\% confidence levels.  It should be noted
that these comparisons ignore the observed Li upper limits prevalent for
the coolest (bin A) Pleiads.  The stars with upper limits lie at the
lower edge of the observed Li abundances (figure 4 of Jones {\it et
al.\/}).  Ignoring this censored data may lead to an {\it underestimate\
\/}of the true dispersion for the coolest Pleiads-- making our
conclusion of no significant difference in the magnitude of star-to-star
abundance scatter for the late G to late K Pleiads a conservative one.
Larger samples and improved upper limits (or detections) would clarify
this important issue.

\section{Nature of the Li-rotation correlation}

\subsection{Projected rotational velocity}
Extant studies of Li-rotation correlations have employed $v \sin i$
measurements, which yield only a lower limit to the rotational velocity
due to the unknown angle of inclination, $i$.  In Figure 5, our
$V-I$-based absolute and differential Li abundances are plotted against
the projected velocity measurement $v \sin i$.  The top two panels (a
and b) show the data for all $T_{\rm eff}$.  The bottom two panels (c
and d) show data with $4500{\leq}T_{\rm eff}{\leq}5500$ K, which is the
temperature range in which a clear connection between Li abundance and
$v \sin i$ was asserted by Garcia Lopez et al.~(1994).  It is seen that
while there is a range of abundances at the lower values of the rotation
velocity, the rapid rotators ($v \sin i{\ga}$ 30 km s$^{-1}$) do show a
tendency to have higher Li abundances in the intermediate $T_{\rm eff}$
range (panel d).
\marginpar{Fig.~5}

\subsection{Rotational Period}
The Pleiades now has many members with photometrically-determined
rotation periods (\markcite{anita98}Krishnamurthi et al.~1998), which
are free of the ambiguity associated with inclination angle.  Hence, it
is now possible to consider the true nature of the correlation between
rotation and Li abundance.  For example, we find a slowly rotating star
(\ion{H}{2} 263, $P=4.8$ d) that has a high Li abundance.  Furthermore,
two of the three stars (\ion{H}{2} 320 and 1124) in the Garcia Lopez et
al.~(1994) study with low $v \sin i$ and high Li abundances also have
measured rotation periods of 4.58d and 6.05d respectively.  Thus, {\it
there are several cases where high Li abundance in stars with low $v
\sin i$ is not due to inclination angle effects\/}-- Li overabundances
are not solely restricted to rapid rotators.  This is apparent in Figure
6, where the surface Li abundance is plotted against rotation period,
P$_{rot}$, instead of $v \sin i$.  In particular, we draw attention to
the large range in abundances seen at longer rotation periods ($>$4.0
days; panels b and d).  Thus there exist at least a few Pleiads which
are true slow rotators, but have high Li abundances.
\marginpar{Fig.~6}

We next examined the proposal by Garcia Lopez et al.~(1994) that there
is a very clear relationship between rotation and log N(Li) for stars
with M~$\sim$0.7-0.9 M$_{\odot}$.  Figure 7 shows the $V-I$-based Li
abundances versus mass for Pleiads with photometrically-measured
rotation rates.  The symbol size is proportional to the rotation period.
When rotational {\it periods} are considered rather than $v \sin i$
measurements, a true range of Li abundances with rotation is seen in the
mass range 0.7-0.9 M$_{\odot}$-- there are genuine slow rotators with
abundances similar to the fast rotators.  Thus, there appears to be a
range of rotation at all abundances.  Hence, P$_{\rm rot}$ is essential
to study the true correlation.
\marginpar{Fig.~7}

\subsection{Structural effects of rotation}
Rapid rotation affects the structure of a star and hence the derived
mass at a given $T_{\rm eff}$ (\cite{endal79}).  The structural effects
of rotation would alter a star's color such that rapidly rotating
objects would be redder, hence perceived as cooler, and thus be assigned
a lower mass.  We therefore examined the abundances as a function of
mass rather than effective temperature.

To investigate this issue, it was necessary to construct stellar models
for different disk lifetimes (\cite{anita97}).  We ran models for
$\omega_{crit}$ = 5$\times \omega_{\odot}$ and 10$\times \omega_{\odot}$
to represent fast rotators and the slow rotators.  The
rotation-corrected masses were derived by interpolation in the models
across effective temperature and rotation velocity for different disk
lifetimes.  We found that the percent change in mass is small (5\%) even
for the most rapidly rotating star in the Pleiades (\ion{H}{2} 1883, $v
\sin i$=140 km s$^{-1}$).  The change is between 1\% and 2\% for stars
with $v \sin i$ in the 50-100 km s$^{-1}$, and $<1$\% for $v \sin i
\leq$ 50 km s$^{-1}$.  These small alterations fail to eliminate the Li
dispersion, which sets in at M$<$0.9 M$_{\odot}$ in the Li-mass
plane. Thus, the structural effects of rotation on the derived
mass-temperature relation are not large enough to account for the
Pleiades Li abundance dispersion.

These results differ with those of Martin \& Claret (1996), who also
explored the structural effects of rotation and found enhanced Li
abundances for stars with high initial angular momentum.  This is not
seen in our models, which predict small rotational structure effects on
the Pleiades Li abundances, similar to the models of Pinsonneault et
al.~(1990).  \markcite{M99}Mendes et al.~(1999) have noted the conflict
between the results of Martin \& Claret (1996) and Pinsonneault et
al.~(1990), and considered the hydrostatic effects of rotation on
stellar structure and Li depletion using their own stellar models.
Their results are in agreement with Pinsonneault et al.~(1990), and they
too find that hydrostatic effects are too small to explain the observed
Li abundance spread in the Pleiades.

\section{Li and Stellar Activity} 

\subsection{Li and chromospheric emission} 
Since the large Pleiades Li spread is in such a young cluster, one may
wonder if its long-sought explanation is related to stellar activity.
Additionally, since rotation and activity are well-correlated, a
Li-activity relation may be masquerading as a Li-rotation relation
instead.  Here, we discuss if magnetic activity indicators such as
chromospheric emission (CE) are correlated with the Li abundance.
Several studies have pointed out that activity is correlated with the Li
abundance (e.g., \cite{s93b}, \cite{jones96}).  There have also been
some studies speculating that CE affects the {\it apparent} abundance of
Li (e.g., \cite{houd95}) and hence may be at least partly responsible
for the dispersion.

Figure 8 contains our results, and shows the $V-I$-based differential
Li abundances versus the \ion{Ca}{2} infrared triplet fluxes (top panel)
and residual fluxes (bottom panel).  A relation is seen in both panels,
such that the lowest log N(Li) differences are seen predominantly for
the lowest flux ratios while the largest log N(Li) differences are seen
predominantly for stars having the largest flux ratios.  The ordinary
correlation coefficients are significant at the 99.7\% and
${\geq}99.9$\% confidence levels for the chromospheric Ca fluxes and
residual fluxes, suggesting a significant relation between chromospheric
activity differences and Li abundance differences (though not
necessarily causal).  
\marginpar{Fig.~8}

\subsection{Spreads in Other Elements}
 
Important clues to the cause of the Pleiades Li abundance scatter 
can be found from examination of other elements not destroyed in stellar 
interiors like $^7$Li.  Variations in such abundances may signal effects 
other than differential Li processing, and perhaps point to an illusory 
difference caused by inadequate treatment of line formation. 
 
\subsubsection{Potassium}
One of the most useful features for this purpose is the ${\lambda}7699$
\ion{K}{1} line.  The usefulness of this feature is two-fold.  First,
there is the similarity in electronic configuration with the \ion{Li}{1}
atom and the fact that this particular K transition and the
${\lambda}6707$ \ion{Li}{1} are both neutral resonance features.
Second, the interplay of abundance and ionization effects leads to the
happy circumstance that the line strengths of these two features are
comparable in Pleiades dwarfs.  Thus, line formation for both features
should be similar in many respects.
 
The \ion{K}{1} line strengths were taken from S93a and Jones {\it et
al.\/}~(1996).  These were then plotted versus $T_{\rm eff}$ as derived
from both $B-V$ and $V-I$.  The relation was well fit by a 4th order
Legendre polynomial approximated by the relation: ${\rm
EW(K)}=9098.151-(4.126458{\times}T)+(6.381173{\times}10^{-4}{\times}T^2)-(
3.308942{\times}10^{-8}{\times}T^3)$ for the $B-V$ colors and by the
relation: ${\rm EW(K)}=8926.171-(4.170605{\times}T)+
(6.702972{\times}10^{-4}{\times}T^2)-(
3.642286{\times}10^{-8}{\times}T^3)$ for the $V-I$ colors.  These fits
showed considerable scatter -- the line strength dispersion was
${\sim}55$ m{\AA}, which is considerably larger (and statistically
significant) than even the maximum equivalent width errors estimated by
S93a.  So scatter is present in the potassium data as well.
 
Differential \ion{K}{1} equivalent widths ( [observed$-$fitted]/fitted )
are plotted against the differential Li abundances in Figure 9.  A
correlation between the values is present, though with considerable
scatter.  The one-sided correlation coefficients are significant at the
99.0 and 98.3\% confidence levels for the $(B-V)$ and $(V-I)$-based
results.  Such a correlation (of whatever magnitude), however, may arise
not from some physical mechanism; instead, it may simply reflect
correlated measurement errors.
\marginpar{Fig.~9}

Like the differential Li abundances, the differential \ion{K}{1} 
line strengths are correlated with activity measure.  Figure 10 shows
the differential \ion{K}{1} equivalent width versus the \ion{Ca}{2} 
fluxes (top panel) and residual fluxes (bottom panel).  The correlations
are analogous to those seen for the differential Li abundances in 
Figure 8, and are significant at the ${\sim}98.5$\% (top panel) and 
${\ge}99.9$\% (bottom panel) confidence levels. 
\marginpar{Fig.~10}

\subsubsection{Calcium}
To examine the possibility of correlated measurement errors, we considered 
the line strengths of the $\lambda$6717 \ion{Ca}{1} feature taken from S93a.  
These were fitted against $T_{\rm eff}$ in the same manner as the \ion{K}{1} 
equivalent widths.  The relations are given by: 
${\rm EW(Ca)}=4203.706-(1.859218{\times}T)+(2.888087{\times}10^{-4}{\times}T^2)
-
(1.538325{\times}10^{-8}{\times}T^3)$ for the $B-V$ colors and by ${\rm
EW(Ca)}=2978.740-(1.249348{\times}T)+(1.887209{\times}10^{-4}{\times}T^2)-
(9.973237{\times}10^{-9}{\times}T^3)$ for the $V-I$ colors.  The scatter
associated with these fits is ${\sim}20$ m{\AA}, which is consistent
with the S93a uncertainties.  Interestingly, unlike Li and K, there is
no evidence for scatter in the calcium data above the measurement
uncertainties.  Differential \ion{Ca}{1} equivalent widths are plotted
against the differential Li abundances in Figure 11.  The relation is
flat.  Unlike K, there is no significant correlation-- the ordinary
correlation coefficients are significant at only the ${\sim}80$\%
confidence level.
\marginpar{Fig.~11}
 
This indicates to us that the correlated scatter in Li and K line
strengths is not due to measurement errors.  Rather, we suggest that
some physical mechanism affecting the details of line formation not
included in standard LTE model photosphere analyses is the cause.  Such
a mechanism, if having star-to-star differences, may be the dominant
source of the Li abundance scatter in Pleiades dwarfs.  Since activity
evinces such differences, it may naturally provide such a mechanism.

In an important theoretical study, Stuik {\it et al.\/}~(1997) have 
urged similar caution in regarding Pleiades Li scatter as solely 
due to genuine abundance differences.  These authors consider the 
photospheric effects of activity on Pleiades \ion{Li}{1} and \ion{K}{1} 
line strengths by modeling surface spots and plages.  They can neither 
exclude nor confirm these particular manifestations of magnetic activity 
as the cause of the problematic and important \ion{K}{1} variations in 
cool Pleiads.  Their extensive efforts, though, do open the door for future 
improvement. 

First, it seems important to establish whether their empirical solar-based
spot/plage models or their ``best-effort'' theoretical stellar models
are more nearly correct, and if one or the other model set is indeed 
applicable to all Pleiads since the two model sets produce color and line 
strength changes opposite in sign.  Second, Stuik {\it et al.\/}~(1997)
note that their radiative equilibrium and mixed activity calculations depart
from observations with increasing $(B-V)$.  As they acknowledge, 
such disparities may signal other effects not yet considered:  UV "line
haze", which may depend on the presence and structure of an overlying 
chromosphere, impacting the details of line formation; unknown properties 
and effects of Pleiads' granulation patterns; and the influence of so-called 
solar-like "abnormal granulation" within plages.  Third, other sources 
of non-thermal heating of the photosphere by chromospheric `activity' 
may need to be considered.  Finally, simply relating colors or effective 
temperatures (from color-$T_{\rm eff}$ conversions) of Pleaids having 
different activity levels may be more problematic than realized. 

Houdebine \& Doyle (1995; HD95) demonstrate that formation of the
$\lambda$6707 \ion{Li}{1} line is sensitive to activity in M dwarfs.
The extent of these effects depends on the relative coverage of plages
and spots.  HD95 note the particular importance of the role of
ionization in reducing the resonance line's optical depth.  In late G
and K dwarfs like those showing scatter in the Pleiades, star-to-star
variations in departures of both photoionisation and collisional
ionization from that predicted by model photospheres might introduce
significant star-to-star variations in the derived Li
abundance\footnote{Overionization from photospheric convective
inhomogeneities has been discussed in the context of Population II star
Li abundances by \markcite{Ku95}Kurucz (1995).}.  Interestingly,
\markcite{Ki99}King et al.~(1999) find element-to-element abundance
differences in two cool ($T_{\rm eff}{\sim}4500$ K) Pleiades dwarfs and
a similarly cool NGC 2264 PMS member which are ionization potential
dependent.  We suggest that current evidence may implicate non-photospheric 
ionization differences as a likely source of star-to-star Li variations 
in the Pleiades. 

\section{Other Mechanisms and Concerns} 
 
\subsection{Metal abundance variations}
Variations in abundances of other elements can affect stellar Li
depletion via the effects of stellar structure on PMS Li burning.  For
example, Figure 3 of Chaboyer, Demarque, \& Pinsonneault (1995),
indicates that very small metal abundance differences of, say, 0.03 dex
lead to substantial (${\ga}0.3-0.4$ dex) differences in PMS Li burning
for $T_{\rm eff}{\la}4500$ K.
 
Extant studies (\cite{BF90}; \cite{C88}) of Pleiades F- and G-star {\it
iron\ \/} abundances (which cannot simply be equated with
``metallicity'' when it comes to PMS Li depletion; \cite{S94}) suggest
no intrinsic scatter larger than 0.06-0.10 dex.  The photometric scatter
of the single stars in the color-magnitude diagram might allow a
metallicity (or, perhaps more properly, those elements which are
dominant electron donors in the stellar photospheres) spread of 0.05
dex.  While small, these constraints would still permit substantial Li
abundance spreads for {\it cool\ \/}Pleiads.  Additionally, abundances
of elements which may have a large impact on PMS Li depletion but little
effect on atmospheric opacity (e.g., oxygen) have yet to be determined
in cool Pleiades dwarfs.
 
However, the Li spread in the Pleiades extends to $T_{\rm eff}$ values
substantially hotter than ${\sim}4500$ K.  At hotter $T_{\rm eff}$
values, model PMS Li-burning is less sensitive to metallicity.  For
example, in the range 5000-5200 K, the observed Li abundance spread
would require ``metallicity'' differences approaching a factor of two.
Such spreads would be surprising indeed, and not expected based on the
limited results of extant Fe analyses of hotter cluster dwarfs.
Abundance differences (of a large number of elements) of this size would
not be difficult to exclude with good quality spectra as part of future
studies.

\subsection{Magnetic Fields} 

\markcite{V98}Ventura et al.~(1998) have recently investigated the
effects of magnetic fields in stellar models and PMS Li depletion.  They
find that even small fields are able to inhibit convection, and thus PMS
Li depletion.  They suggest that a dynamo generated magnetic field
linked to rotational velocity (thus, presumably yielding an association
between activity and rotation given conventional wisdom) would result in
ZAMS star-to-star Li variations that mirror differences in star-to-star
rotational (and presumably activity) differences.  As these authors
admit, the fits of their magnetic models to the Li-$T_{\rm eff}$
morphology and significant scatter of the Pleiades observations are not
``perfect'' or ``definitive''; however, the qualitative agreement and
ability to produce star-to-star scatter and general relations between Li
abundance and rotation and activity are encouraging.  Continued
observations (especially detailed spectroscopic abundance determinations
of various elements in numerous Pleiads) and theoretical work will be
needed to establish the degree to which the Pleiades Li spread is
illusory or real and, if the latter, its cause(s).

\section{Summary and Conclusions}

The very large dispersion in Li abundances at fixed $T_{\rm eff}$ in
cool ($T_{\rm}{\la}5400$ K) Pleiads is a fundamental challenge for
stellar evolution because standard stellar models of uniform age and
abundance are unable to reproduce it.  A variety of mechanisms
(rotation, activity, magnetic fields, and incomplete knowledge of line
formation) have been proposed to account for this scatter.  Here, we
construct a sample of likely single Pleiads and consider this problem
with: {\ }a) differential Li abundances relative to a mean $T_{\rm eff}$
trend {\ }b) rotational periods instead of projected rotational
velocities {\ }c) chromospheric emission indicators, and {\ }d) line
strengths of other elements.

We calculated $T_{\rm eff}$ values from both $(B-V)$ and $(V-I)$ on a
self-consistent scale based on the calibrations of Bessell (1979).  We
find differences in the two $T_{\rm eff}$ values and these are
significantly correlated with both general activity level and with
differences in activity, suggesting that surface inhomogeneities may
noticeably affect stellar colors.  Our results are consistent with a
growing body of evidence of significant differences between $(B-V)$- and
$(V-I)$-based $T_{\rm eff}$ values, a propensity for $(B-V)$ to yield
larger $T_{\rm eff}$ values, and a relation of these characteristics
with activity in young stars from 5 Myr old PMS stars in NGC 2264
(Soderblom et al.~1999) to PMS stars in the 30 Myr old IC 2602 (Randich
et al.~1997) to ZAMS stars in the ${\sim}100$ Myr old Pleiades.

However, the similarity between the sensitivity of the derived Li
abundance to $T_{\rm eff}$ and the clusters' physical Li-$T_{\rm eff}$
morphology means that even substantial $T_{\rm eff}$ errors are not a
significant source of star-to-star Li scatter.  Nor are observational
errors.  Comparison of the scatter in the differential Li abundances
with errors from $T_{\rm eff}$ and line strength uncertainties indicates
an infinitesimal probability that the observed scatter occurs by chance.
We find significant scatter in the Li abundances below ${\sim}6000$ K;
it is significantly larger, though, below ${\sim}5500$ K.  Statistical
analysis fails to support previous claims of smaller scatter in the late
K dwarfs relative to the late G and early-mid K dwarfs.

There is a spread of Li abundance at low $v \sin i$, whereas the rapid
{\it projected\ \/}rotators tend to have larger differential Li
abundances in the range $4500{\leq}T_{\rm eff}{\leq}5500$.  However, use
of photometric rotation periods (free from uncertainties in the
inclination angle $i$) indicates there is {\it not\ \/}a one-to-one
mapping between differential Li abundance and rotation.  The stars
\ion{H}{2} 263, 320, and 1124 are examples of stars with Li excesses but
slow rotation (P$=4.6-6.1$ d).  In contrast to previous claims based on
$v \sin i$, the rotation periods indicate a true range of Li abundance
with rotation in the mass bin 0.7-0.9 M$_{\odot}$.

Using the theoretical framework of Krishnamurthi et al.~(1997), we
constructed stellar models to investigate the hydrostatic effects of
rotation on stellar structure and PMS Li burning.  As shown in Figure 8,
these models fail to account for the Pleiades Li dispersion, which is in
agreement with the independent conclusions of Mendes et al.~(1999).

We find that the star-to-star differences in Pleiades Li abundances are
correlated with activity differences, as measured from \ion{Ca}{2}
infrared triplet flux ratios, at a statistically significant level.
Moreover, the Li differences are significantly correlated with
differences in the strengths of the $\lambda$7699 \ion{K}{1} resonance
feature.  This seems to not be due to correlated measurement errors
since the Li differences show no correlation with the $\lambda$6717
\ion{Ca}{1} line strength residuals.  This is a significant result given
similarities in the Li and K feature's atomic properties and line
strengths.  We suggest that incomplete treatment of line formation,
related to activity differences, plays a significant role in the Li
dispersion-- i.e., that part of the dispersion is illusory.  As
emphasized by Houdebine \& Doyle (1995), the formation of the
\ion{Li}{1} feature is sensitive to ionization conditions.  If
chromospheric activity variations can produce significant variations in
photo- and collisional-ionization in the \ion{Li}{1} line formation
region not accounted for by LTE analyses using model photospheres, this
may lead to errors in the inferred abundance.  Relatedly, we note the
results of King {\it et al.\/}~(1999) who found ionization
potential-dependent effects in the elemental abundances of two cool
($T_{\rm eff}$) Pleiads and a similarly cool NGC 2264 PMS star.

If such conjecture is correct, we expect that somewhat older (less
active) cluster stars will exhibit less Li dispersion.  This seems to be
the case for the ${\sim}800$ Myr old Hyades cluster
(\markcite{T93}Thorburn et al.~1993) and perhaps also for the mid-G to
mid-K stars in M34 (Jones et al.~1997).  These clusters still exhibit
scatter, and this may be real and due to differences in depletion from
structural effects of rotation (Mendes et al.~1999), magnetic fields
(Ventura et al.~1998), small metallicity variations (\S5.1), main
sequence depletion due to angular momentum transport from spin-down
(Pinsonneault et al.~1990; \markcite{Ch92}Charbonnel et al.~1992) or a
planetary companion (\markcite{Co97}Cochran et al.~1997), and
photospheric accretion of circumstellar/planetary material
(\markcite{Al67}Alexander 1967; \markcite{Go98}Gonzalez 1998).  The
amount of scatter expected in even older clusters is less clear.  If,
e.g., rotationally-induced mixing acts over longer timescales then the
scatter may well increase again; indeed, the substantial Li scatter in
M67 solar-type stars observed by \markcite{Jo99}Jones et al.~(1999)
could indicate that this is the case.

These authors called attention to the possible pattern of very large Li
scatter in young clusters, considerably reduced scatter in
intermediate-age clusters, and increased scatter in older clusters.  In
the scenario we envision, variations in activity-regulated ionization of
the \ion{Li}{1} atom may be responsible for the majority of (mostly
illusory) star-to-star Li differences in near-ZAMS and younger stars; of
course, this does not exclude a (lesser) role from other variable
mechanisms influencing PMS Li burning.  If the decline in the activity
level of intermediate-age stars reduces the importance of variable
ionization, then the (smaller) Li scatter in these stars could arise
from variations in PMS Li burning due to, e.g., the hydrostatic effects
of rotation on stellar structure, inhibition of convection by magnetic
fields, and small metallicity variations; additional contributions may
come from processes just beginning to become effective for ZAMS stars
such as rotationally-induced mixing and planetary/circumstellar
accretion.  In older stars such as M67, the increase in scatter (and
overall Li depletion) is then a product of processes efficiently acting
on the main-sequence proper such as rotationally-induced mixing and/or
photospheric accretion.

Distinguishing specific mechanisms and their relative importance for Li
depletion and scatter at a given age will require continuing
observational and theoretical efforts.  Important advances on the
theoretical front are at least three-fold: {\ }continued investigation
of the role of magnetic fields in PMS Li depletion, realistic model
atmospheres which include chromospheres, and detailed NLTE abundance
calculations which employ these to extend extant sophisticated 
modeling attempts (e.g., Stuik {\it et al.\/}~1997).  On the observational 
front, continued observations of Li in a variety of clusters spanning a range
in age and metallicity are needed.  We believe that particularly
important observational work to be accomplished includes the
determination of photometric periods in more cluster stars, detailed
abundances of numerous elements (in particular, using both ionization
sensitive and ionization insensitive features and elements) in cluster
stars, quantification of even small ``metal'' (not just Fe) abundance
spreads in cluster stars, and the association between planetary systems
and parent star Li and light metal abundances.

\acknowledgements
AK acknowledges support from NASA grant H-04630D to the University of 
Colorado.

\clearpage


\begin{table}
\begin{center}
\epsfxsize=0.9\hsize
\epsfbox{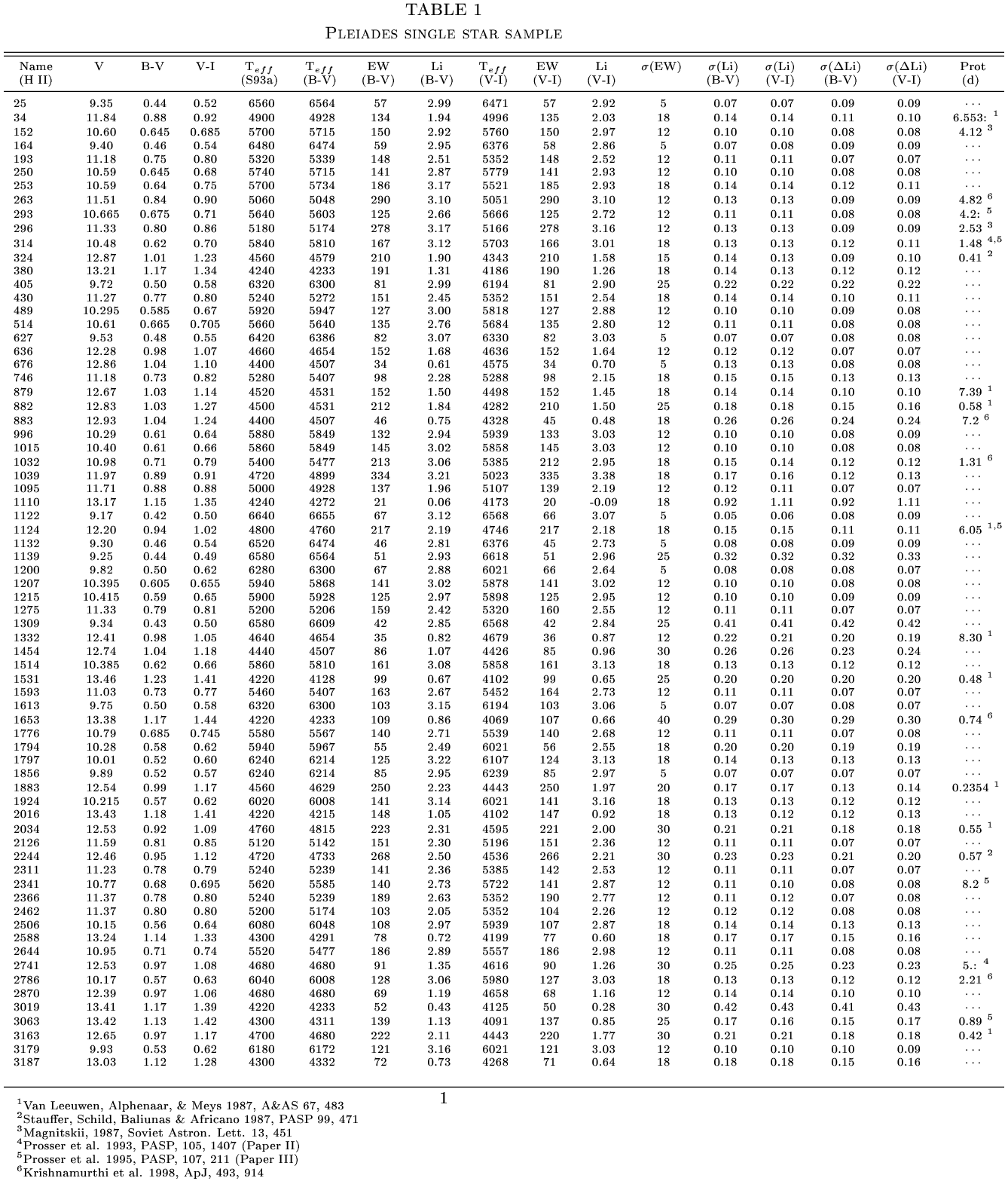}
\end{center}
\end{table}


\begin{figure} 
\plotone{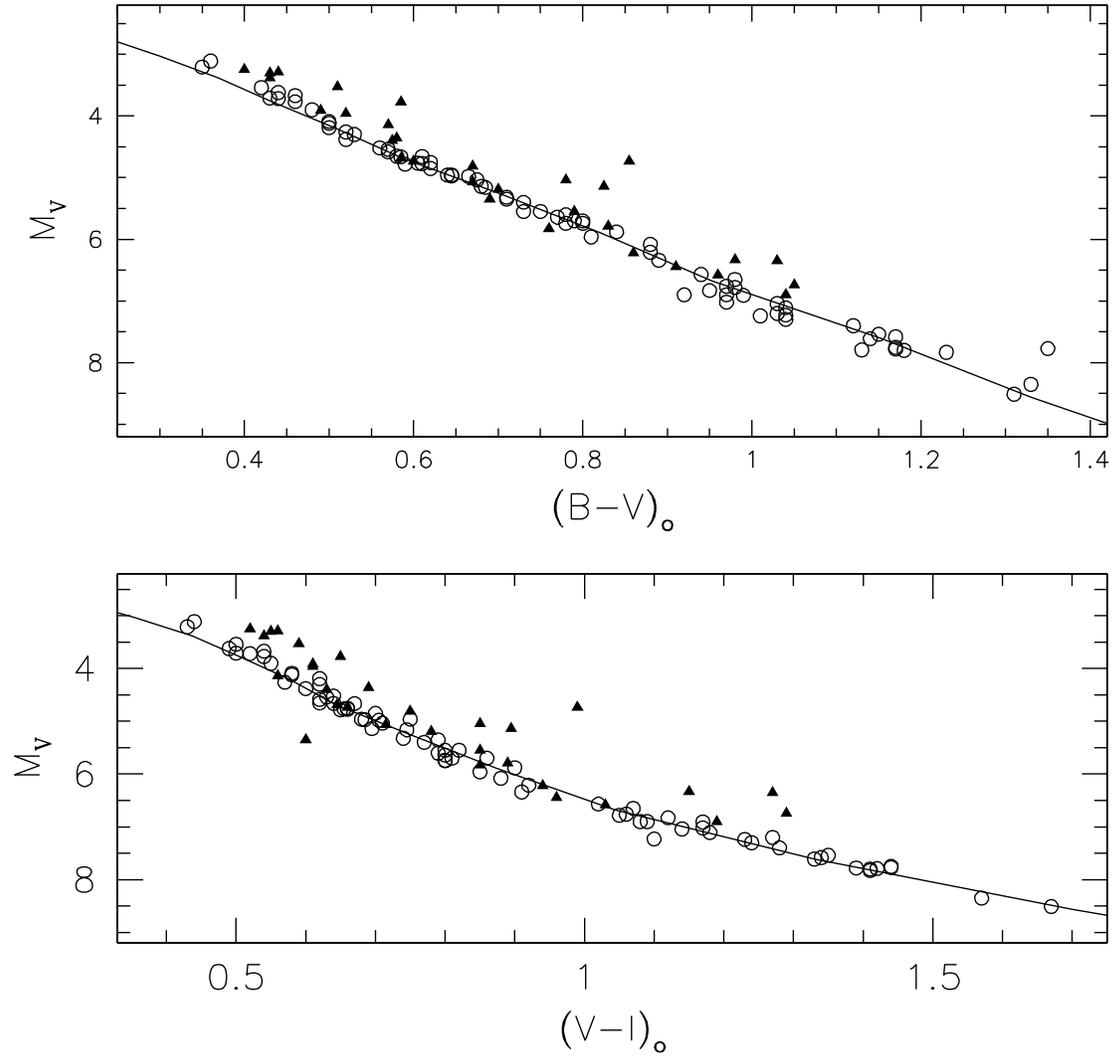}
\caption[]{The color magnitude diagrams of our final non-binary Pleiades
Li sample (open circles) and stars rejected as binaries (filled
triangles).  The Pleiades data is plotted assuming a distance modulus of
5.63 and with a 100 Myr isochrone from Pinsonneault {\it et
al.}~(1998).} 
\end{figure}

\begin{figure} 
\plotone{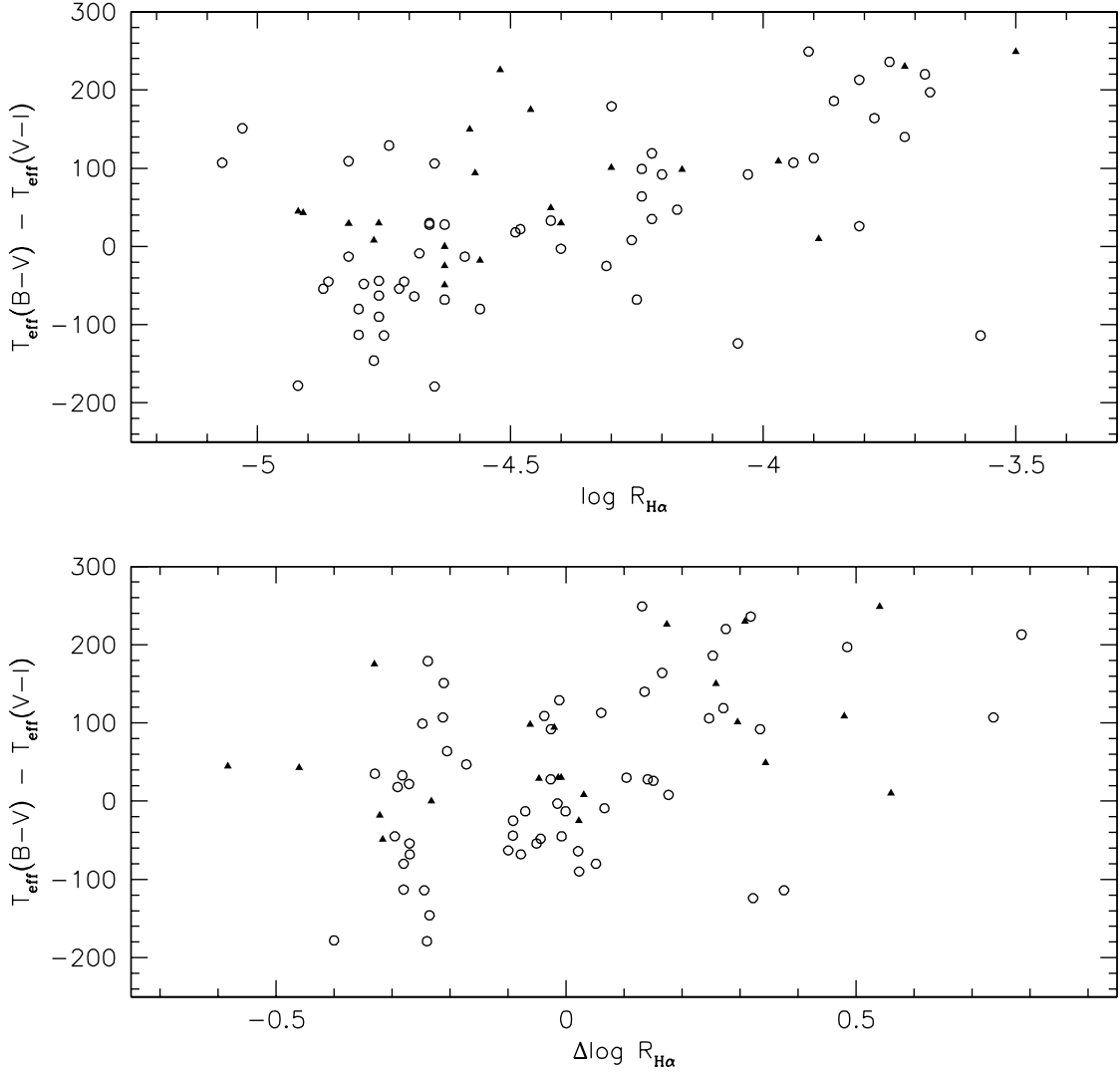}
\caption[]{The top panel shows the difference between our $(B-V)$- and  
$(V-I)$-based $T_{\rm eff}$ estimates versus H$\alpha$ flux ratio (relative 
to the stellar bolometric flux) from Soderblom {\it et al.\/}~(1993).  The 
bottom panel shows the temperature difference versus the residual H$\alpha$ 
flux ratio, which is the flux ratio less a fitted color-dependency.} 
\end{figure}  

\begin{figure}
\plotone{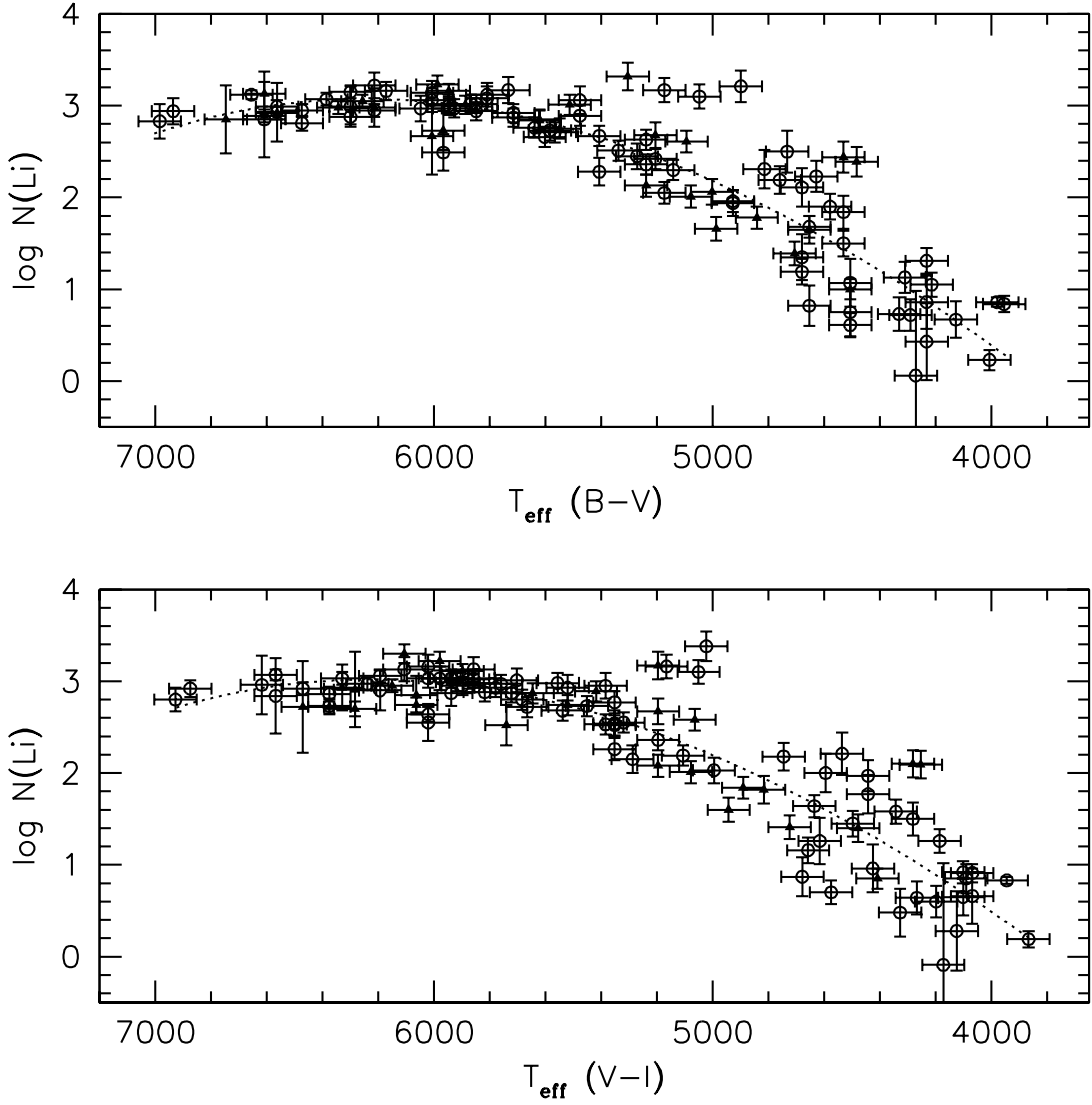}
\caption[]{Li abundances (with derived errors) vs. $T_{\rm eff}$ from our
$(B-V)$ measures (top) and $(V-I)$ measures. The well-known declining trend
of Li abundance with decreasing $T_{\rm eff}$ is fit with a fourth order
Legendre polynomial (dashed line).} 
\end{figure}

\begin{figure}
\plotone{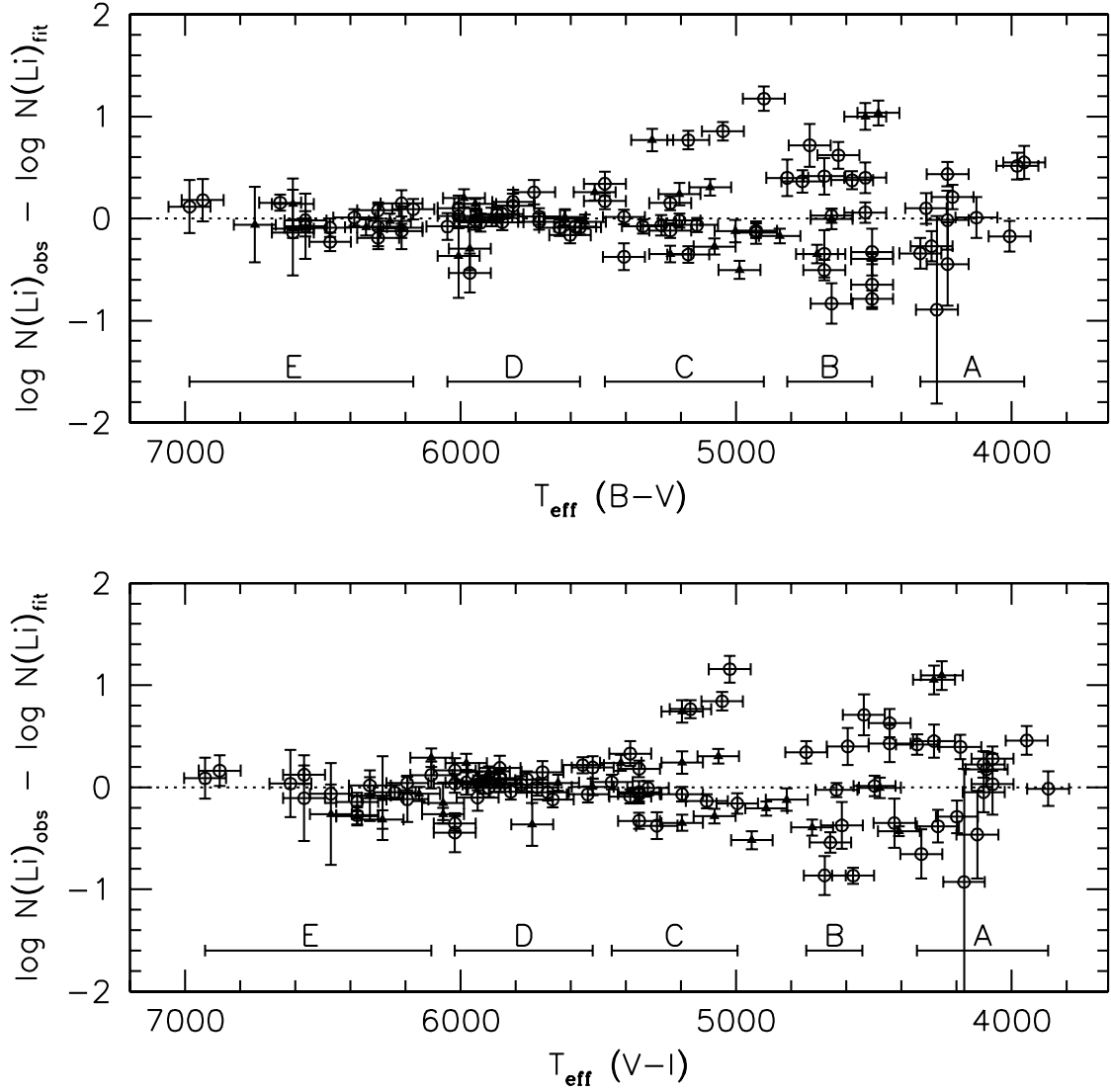}
\caption[]{Same as Figure 3 except the differential (detrended) Li abundances 
and related errors are shown.  Temperature bins used in considering the
scatter of Li abundances as a function of $T_{\rm eff}$ are labeled in
both plots.}
\end{figure}

\begin{figure}
\plotone{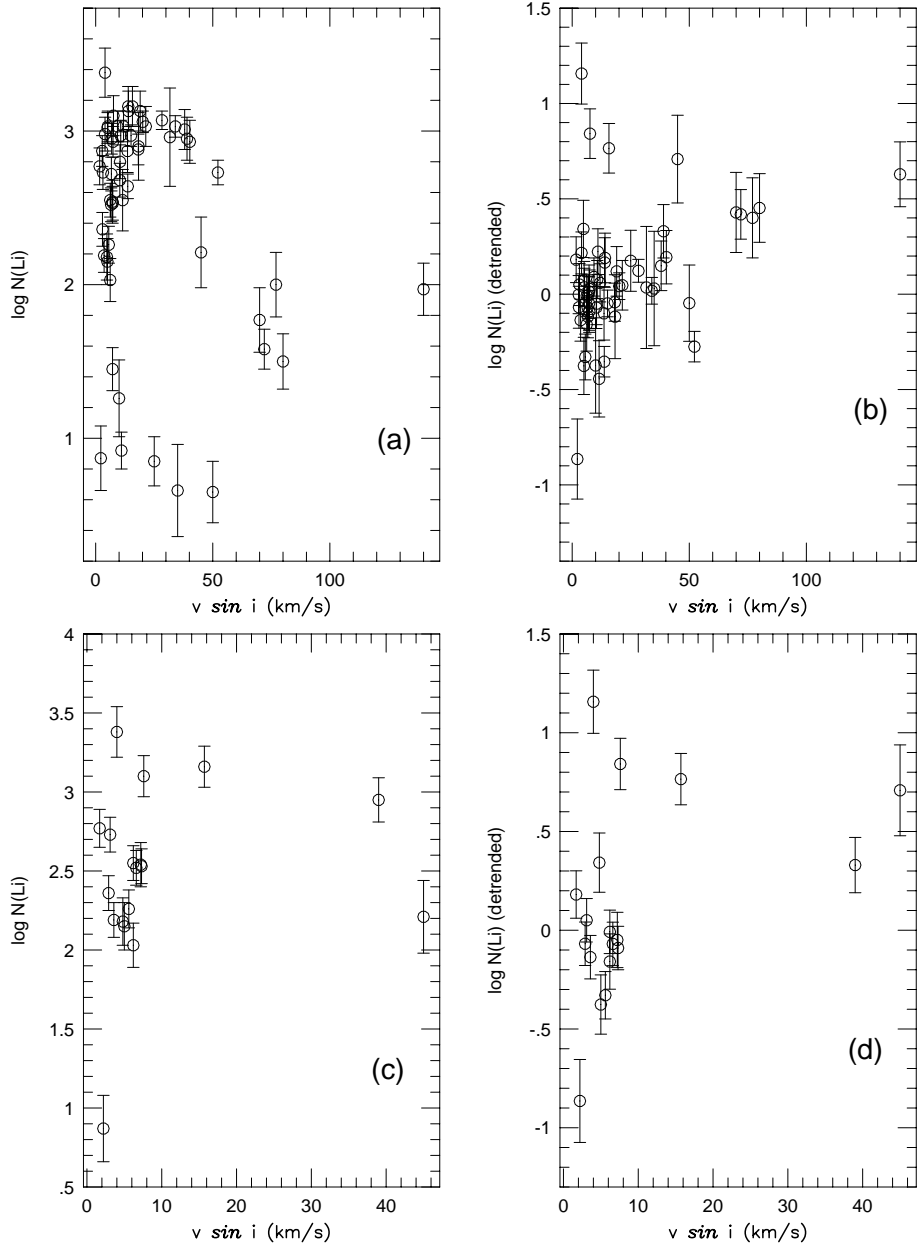}
\caption[]{Projected rotational velocity $v \sin i$ versus $(V-I)$-based
Li results for (a) all $T_{\rm eff}$ and absolute Li abundance (b) all
$T_{\rm eff}$ and differential detrended Li abundance (c)
$T_{\rm eff}=4500-5500$ K and absolute Li abundance (d) $T_{\rm
eff}=4500-5500$ K and differential Li abundance.}
\end{figure}

\begin{figure}
\plotone{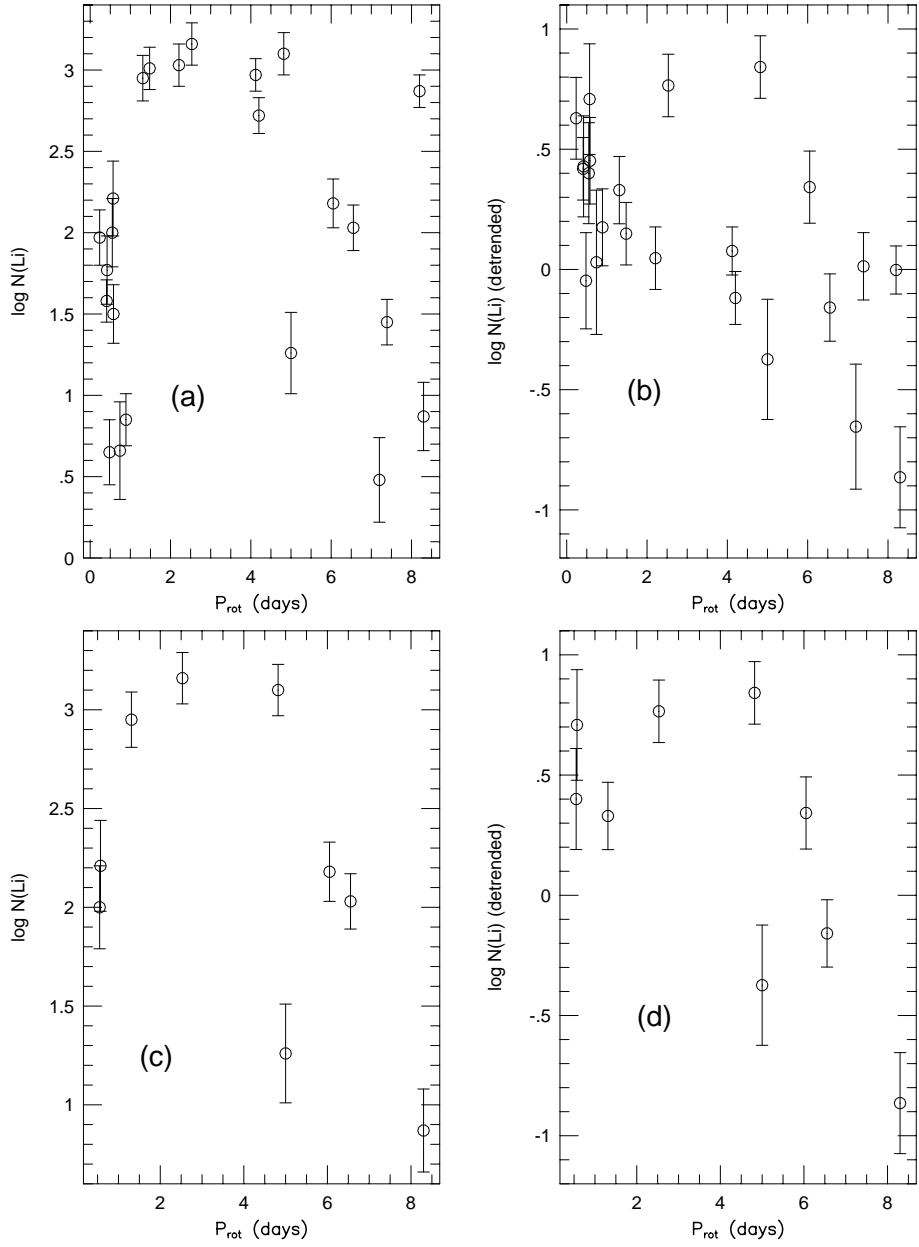}
\caption[]{Same as Figure 5 except rotational period, P$_{rot}$, in days
is plotted instead of $v \sin i$.}
\end{figure}

\begin{figure}
\plotone{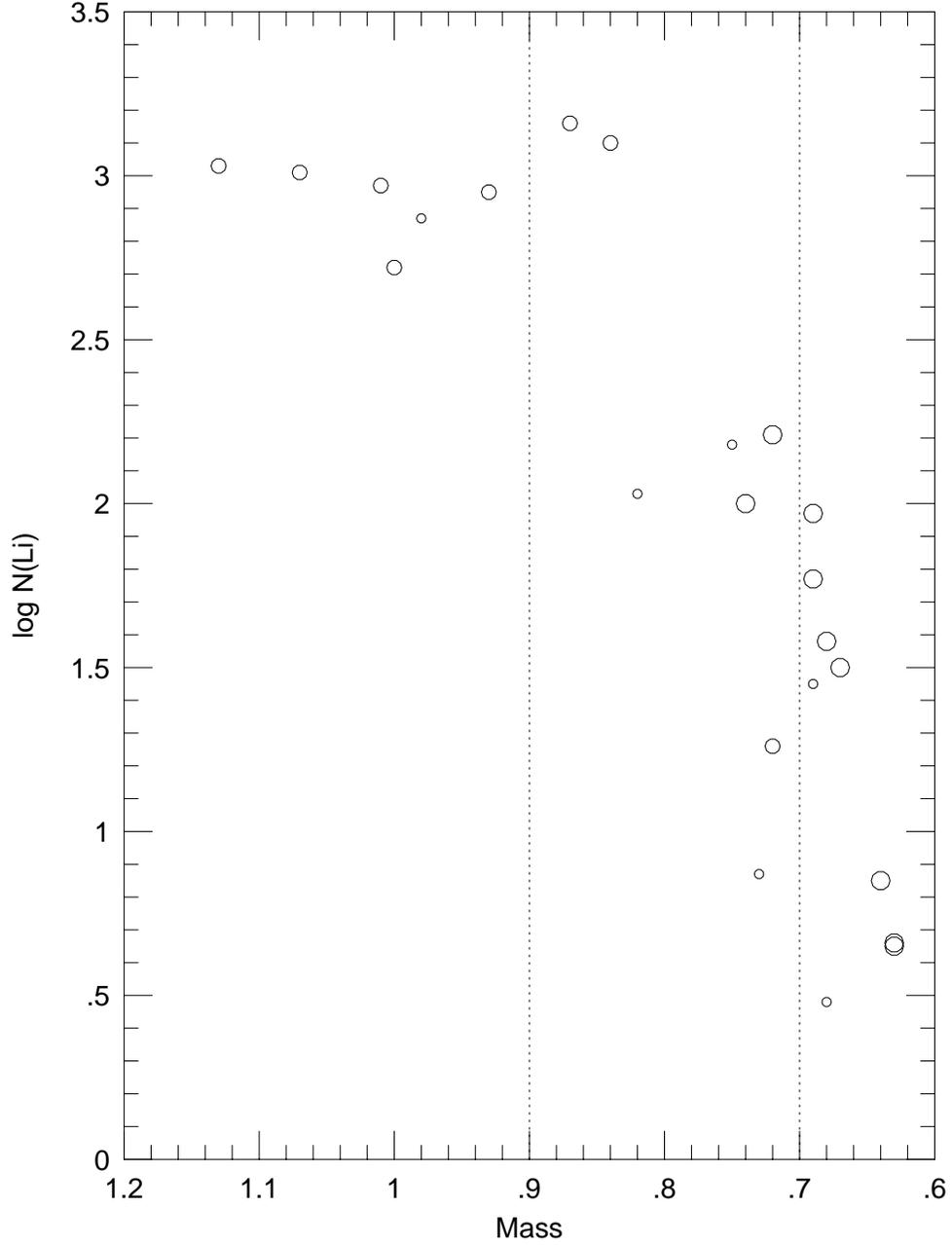}
\caption[]{Our $(V-I)$-based Li abundances versus stellar mass.  The
symbol size corresponds to rotational period.  The largest circles
represent stars with $P_{\rm rot} <$ 1d, the medium sized circles
represent those with $P_{\rm rot}$ between 1d and 5d, and the smaller
circles indicate stars with $P_{\rm rot}$ between 5d and 10d.  The
dashed lines denote the mass range $M=0.7-0.9$ M$_{\odot}$.}
\end{figure}

\begin{figure}
\plotone{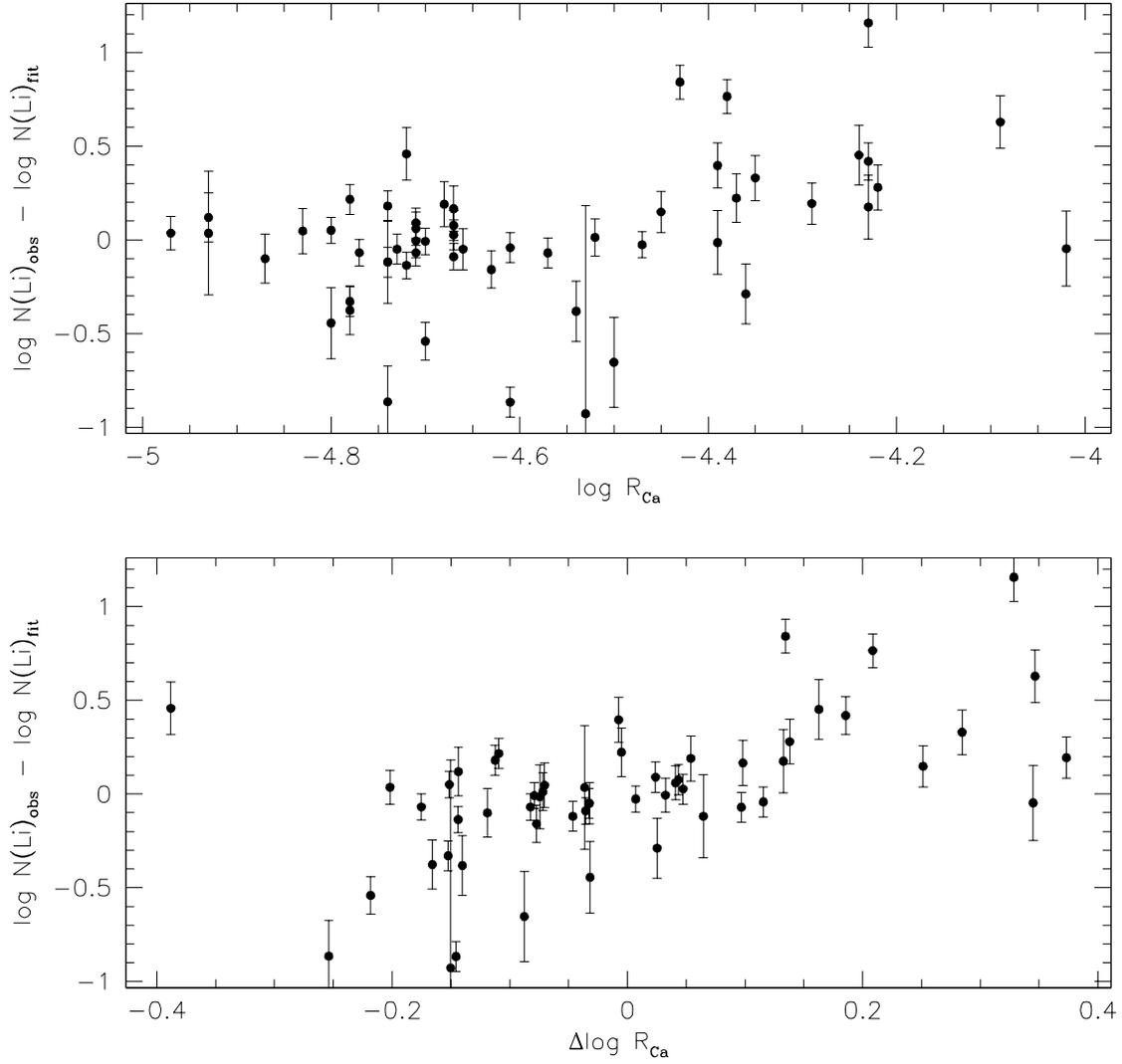}
\caption[]{Our $(V-I)$-based differential Li abundances are plotted
versus the \ion{Ca}{2} infrared triplet flux ratio (top panel) and
residual flux ratio (bottom), which is the flux ratio less a 
fitted color-dependency.}
\end{figure}

\begin{figure}
\plotone{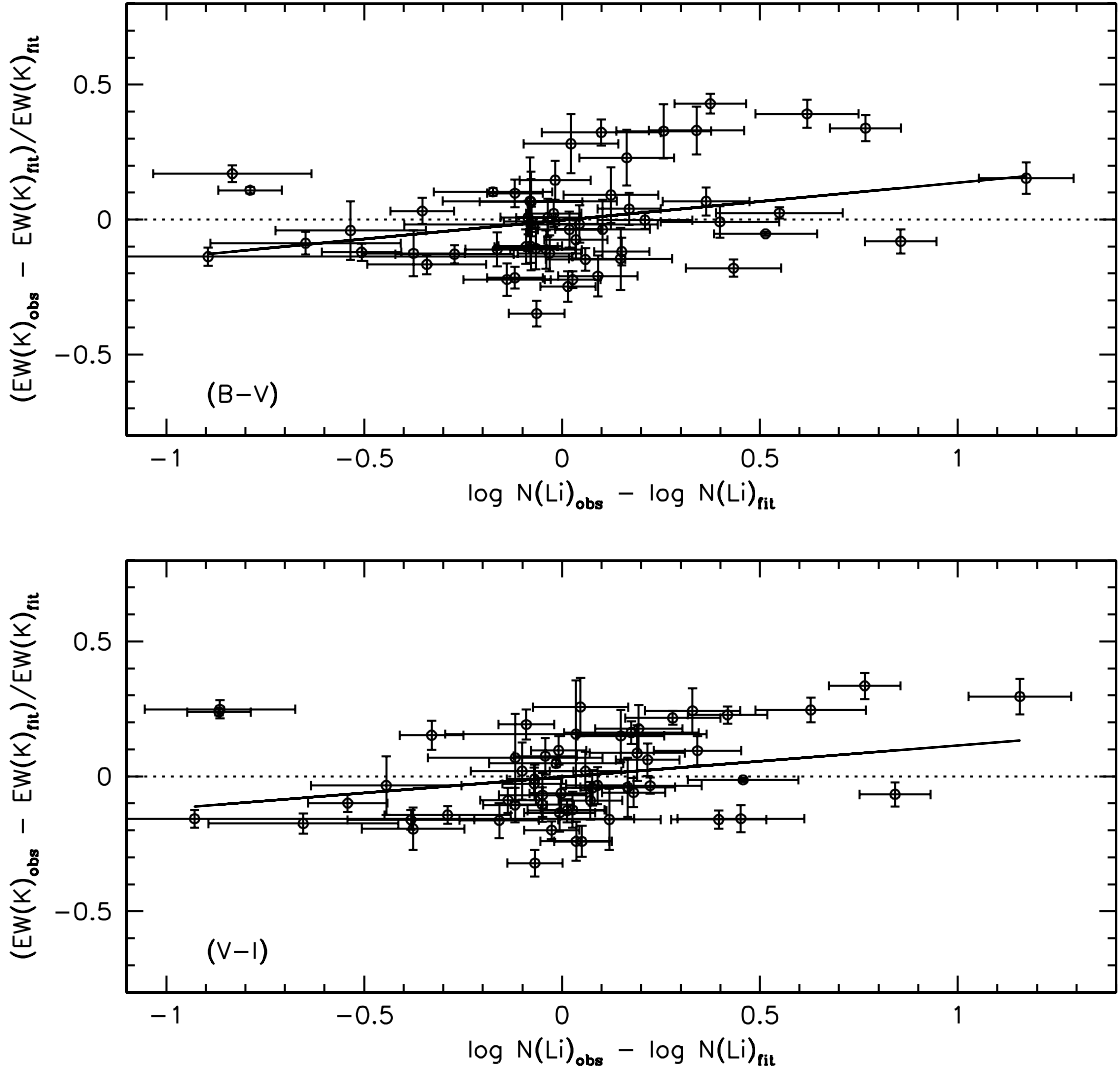}
\caption[]{Differential ${\lambda}7699$ \ion{K}{1} equivalent widths 
([observed$-$fitted]/fitted ) versus our $(B-V)$-based (top panel) and 
$(V-I)$-based differential Li abundances (bottom panel).} 
\end{figure}

\begin{figure}
\plotone{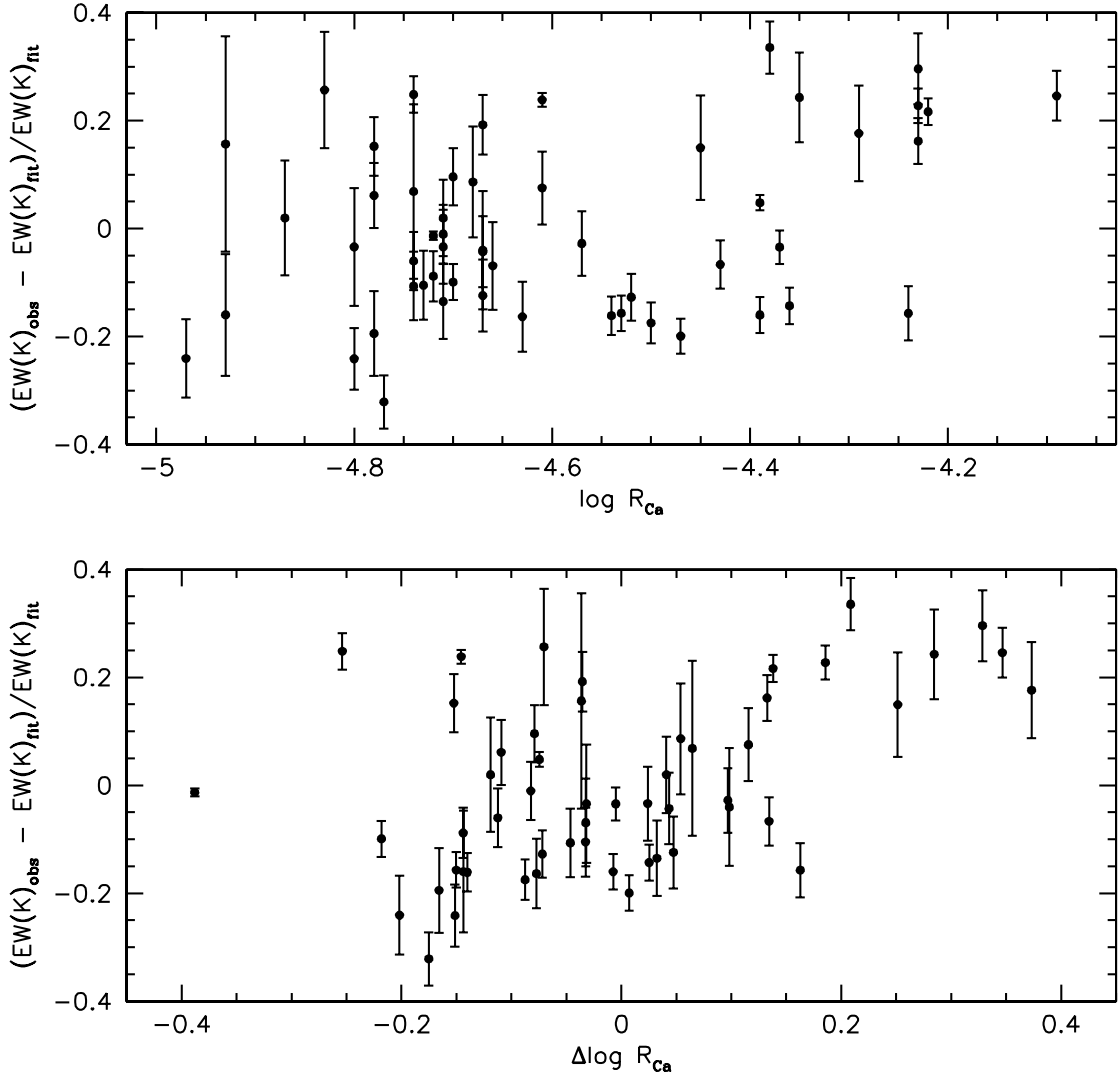}
\caption[]{The $(V-I)$-based differential \ion{K}{1} line strengths are
plotted versus the \ion{Ca}{2} infrared triplet flux ratio (top panel) 
and residual flux ratio (bottom panel).}  
\end{figure}

\begin{figure}
\plotone{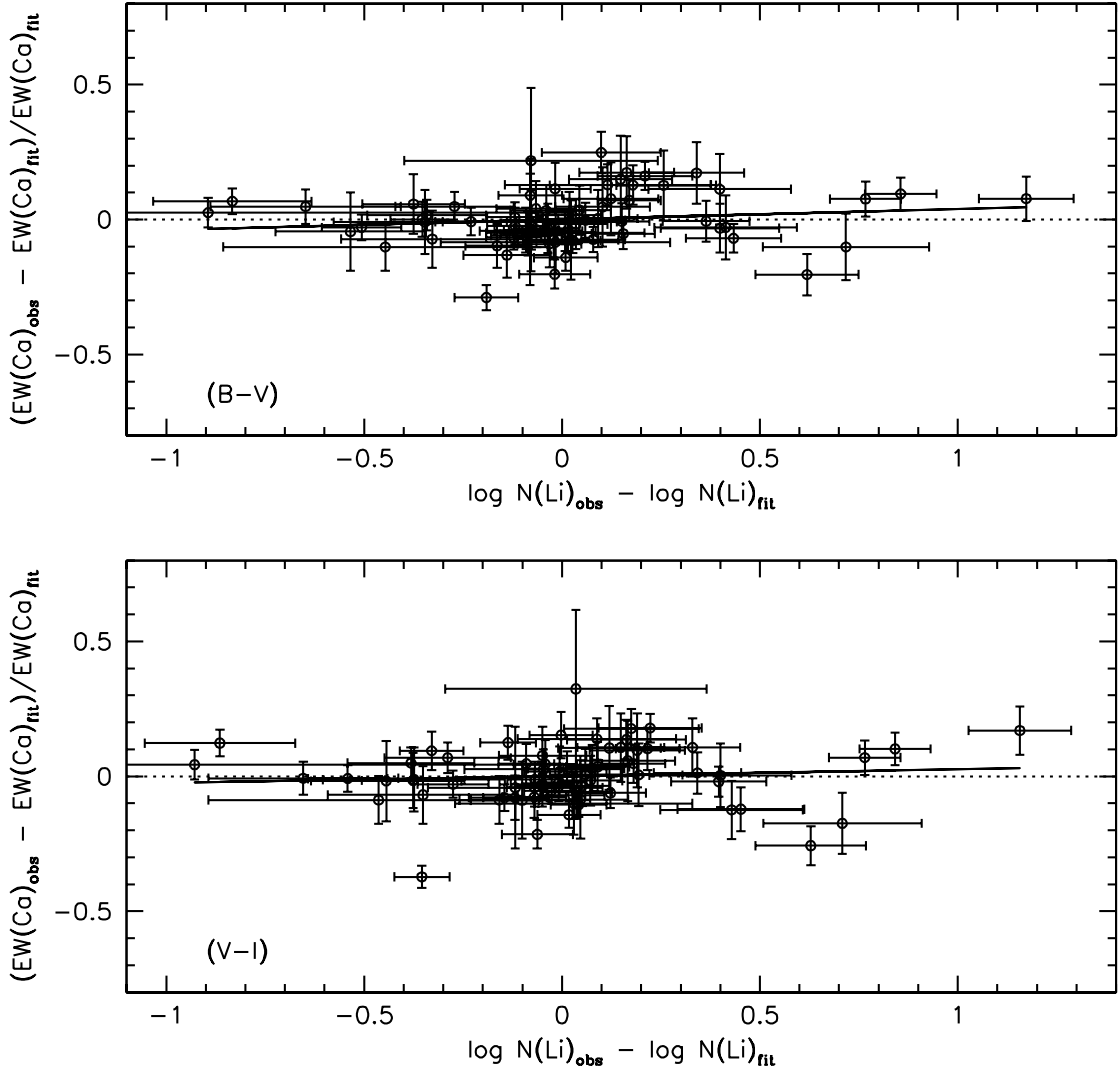}
\caption[]{Differential ${\lambda}6717$ \ion{Ca}{1} equivalent widths 
versus our $(B-V)$-based (top panel) and $(V-I)$-based (bottom panel)  
differential Li abundances.}
\end{figure}

\end{document}